\documentclass[aps, prb, groupedaddress, showpacs, showkeys, twocolumn, 
]{revtex4}
\usepackage{graphics, amsfonts, amsmath, mathrsfs, amssymb, hyperref}
\def\preprint
#1{\thispagestyle{empty}~\newline\vspace*{-22.65mm}
\begin{flushright}
\begin{tabular}{l} #1
\end{tabular}
\end{flushright}
\vspace{1cm}}

\begin{document}
\title{
Qualitative study of perfect-fluid Friedmann-Lema{\^{\i}}tre-Robertson-Walker models with a cosmological constant}

\author{Sebastiano Sonego}
\email{\tt sebastiano.sonego@uniud.it} 
\affiliation{DCFA, Sezione di Fisica e Matematica, Universit\`a di Udine, Via delle Scienze 208, 33100 Udine, Italy}

\author{Vittorino Talamini}
\email{\tt vittorino.talamini@uniud.it}
\affiliation{DCFA, Sezione di Fisica e Matematica, Universit\`a di Udine, Via delle Scienze 208, 33100 Udine, Italy}
\affiliation{INFN, Sezione di Trieste}
\begin{abstract}
The evolution of spatially homogeneous and isotropic cosmological models containing a perfect fluid with equation of state $p=w\rho$ and a cosmological constant $\Lambda$ is investigated for arbitrary combinations of $w$ and $\Lambda$, using standard qualitative analysis borrowed from classical mechanics.  This approach allows one to consider a large variety of situations, appreciating similarities and differences between models, without solving the Friedmann equation, and is suitable for an elementary course in cosmology.
\end{abstract}

\thispagestyle{empty}
\maketitle

\newpage


\section{Introduction}
\label{sec1}
\setcounter{equation}{0}

In introductory courses on cosmology, it is common to study the evolution of the scale factor assuming that the universe is filled with a perfect fluid with equation of state
\begin{equation}
p=w\rho,
\label{w}
\end{equation}
where $p$ is the pressure, $\rho$ is the energy density (including, of course, the density of mass multiplied by $c^2$) and  $w$ is a dimensionless constant.  This covers, in particular, the very important cases of dust ($w=0$) and radiation ($w=1/3$).

Although it is relatively easy to write down the relevant equations (see Sec.~\ref{sec2} below), the time evolution is usually discussed only for a few specific cases.  Often, textbooks present the exact solutions corresponding to dust and radiation, and sometimes even stiff matter ($w=1$), in the absence of a cosmological constant $\Lambda$.  This, however, leads some students to wonder how general are the results so obtained.  In the literature, one can also find analytic solutions for an arbitrary value of $w$ (see, e.g., Refs.~\onlinecite{assad-lima, faraoni, lima}), but a detailed analysis of the properties of such solutions is somewhat lengthy and time-consuming --- certainly not appropriate for introductory lectures.  Furthermore, it must be noted that it is not terribly interesting to know the analytic solutions for simple cases with a definite $w$, given that none of them applies to the entire history of the universe, and that an exact solution in which the universe in different eras is dominated by different components is definitely out of reach.  

Even leaving such considerations aside, the exact solutions that one can find in the literature usually do not cover the possibility of a non-vanishing cosmological constant (see Refs.~\onlinecite{felten, d'Inverno, harvey} for notable exceptions, which are, however, restricted to the case of dust).  Yet the current observational data not only do not exclude, but appear instead to support a cosmological constant~\cite{carroll, earman, lambda} (although other possibilities, involving the broad concept of ``dark energy'', are often considered~\cite{Linder, harvey-lambda}).  Apart from very special cases, finding exact solutions even for a specific value of $w$ and a non-zero $\Lambda$ is usually a mathematical problem well beyond the capability of most of the students, and would anyway take time that is better  used to present other topics of greater physical interest.

One of us was facing these problems while teaching a course whose aim was to give a broad, yet accurate presentation, in 28 hours, of modern physical cosmology to undergraduate students majoring in mathematics, computer science, and engineering.  His strategy was to entirely avoid exact solutions, replacing them by a qualitative study.  Remarkably, such  qualitative analysis, together with the asymptotic behaviour of the scale factor in regimes where it is very easy to solve the equations, allows one to gather most of the relevant information without dealing with unnecessary mathematical technicalities.  Although it is traditional, in elementary courses,  to cover only the range $0\leqslant w\leqslant 1$, this method can be easily applied for all values of $w$.  A paper published a few years ago~\cite{nemiroff} contains a nice discussion of the physical properties of matter, and of their cosmological relevance and implications, with $w$ in different ranges.  Here, we present a complementary analysis, showing how such kinds of matter affect the evolution of the universe at large scales.

Qualitative techniques are well-known and employed in cosmology.~\cite{ellis, uzan, coley, wainwright}  However, no elementary discussion, suitable for being used in an introductory course, is available to our knowledge.~\cite{fn1}  In this paper we give a very simple presentation, based on the often noted analogy between the scale factor and the coordinate of a one-dimensional mechanical system.  Surprisingly, we could not find such a treatment in a fairly large sample of books on cosmology that we have analysed, so we believe it is worth bringing it to the attention of the community of physics teachers.

In Sec.~\ref{sec2} we write down the basic equations.  The mechanical analogue, and an outline of the qualitative analysis, are discussed in Sec.~\ref{sec3}, while Sec.~\ref{sec4} contains the detailed study for physically interesting combinations of $w$ and $\Lambda$.  In Sec.~\ref{sec5} we present some final comments and possible extensions of our approach.  


\section{Equations}
\label{sec2}
\setcounter{equation}{0}

Our notations and settings are standard.  We consider a Friedmann-Lema{\^{\i}}tre-Robertson-Walker (FLRW) model, whose $t=\mbox{const}$ spatial sections are homogeneous and isotropic.  The spacetime metric is 
\begin{equation}
\mbox{\sl g}=-c^2\,\mathrm{d} t^2+a(t)^2\left(\frac{\mathrm{d} r^2}{1-k r^2}+r^2\left(\mathrm{d}\theta^2+\sin^2\theta\,\mathrm{d}\varphi^2\right)\right),
\label{metric}
\end{equation}
where $a(t)>0$ is the scale factor, $k=-1$, $0$, or $1$ according to whether the spatial geometry is hyperbolic, flat, or spherical, and $r$, $\theta$, $\varphi$ are comoving coordinates.  Note that $r$ is dimensionless.  We shall restrict our analysis to the case in which the energy density $\rho$ is strictly positive; an extension to $\rho\leqslant 0$ is mathematically straightforward.

The equations describing the evolution of this model in the presence of a perfect fluid and a cosmological constant $\Lambda$ are:
\vskip .2truecm
\noindent $\bullet$\hskip .3truecm The Friedmann equation
\begin{equation}
\left(\frac{\dot{a}}{a}\right)^2=\frac{8\pi G}{3 c^2}\,\rho+\frac{\Lambda c^2}{3}-\frac{kc^2}{a^2},
\label{Friedmann}
\end{equation}
where, as usual, a dot denotes the derivative with respect to $t$;
\vskip .15truecm
\noindent $\bullet$\hskip .3truecm The equation for the conservation of energy during the cosmological expansion, 
\begin{equation}
\dot{\rho}+3\,\frac{\dot{a}}{a}\left(\rho+p\right)=0;
\label{energy}
\end{equation}
\vskip .15truecm
\noindent $\bullet$\hskip .3truecm A barotropic~\cite{baro} equation of state
\begin{equation}
p=p(\rho),
\label{state}
\end{equation}
usually chosen of the type~\eqref{w}.

Overall, we have three equations for the three unknown functions $a(t)$, $\rho(t)$, and $p(t)$.  Replacing the equation of state~\eqref{w} into Eq.~\eqref{energy} one finds 
\begin{equation}
\frac{\dot{\rho}}{\rho}=-3\left(1+w\right)\frac{\dot{a}}{a},
\label{rho-a-diff}
\end{equation}
which can be immediately integrated to obtain
\begin{equation}
\rho=\frac{\rho_0\, a_0^{3(1+w)}}{a^{3(1+w)}},
\label{rho-a}
\end{equation}
where the subscript 0 corresponds to some arbitrary reference time $t_0$.  This can be replaced into the Friedmann equation~\eqref{Friedmann} to get a single differential equation for the unknown function $a(t)$:
\begin{equation}
\frac{\dot{a}^2}{2}-\frac{4\pi G\rho_0\,a_0^{3(1+w)}}{3 c^2}\,\frac{1}{a^{1+3w}}-\frac{\Lambda c^2}{6}\,a^2=-\frac{k c^2}{2},
\label{Friedmann'}
\end{equation}
where we have multiplied by $a^2/2$ and rearranged the terms for later convenience.


\section{Equivalent mechanical system}
\label{sec3}
\setcounter{equation}{0}

\begin{figure}[htbp]
\vbox{ \hfil
\scalebox{0.8}{
{\includegraphics{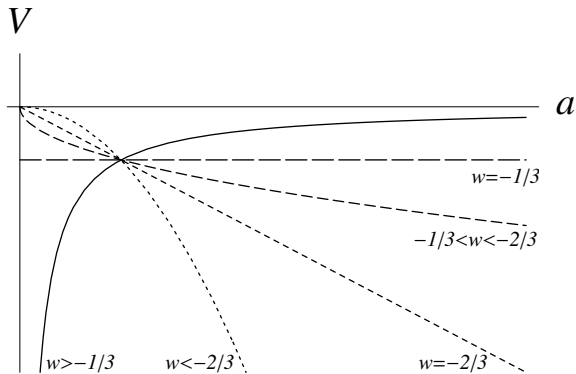}}}
\hfil }
\vskip.25cm
\caption{\small The contribution $-A/a^{1+3w}$ to the potential energy for different values of $w$.}
\label{F:1a}
\end{figure}

\begin{figure}[htbp]
\vbox{ \hfil
\scalebox{0.8}{
{\includegraphics{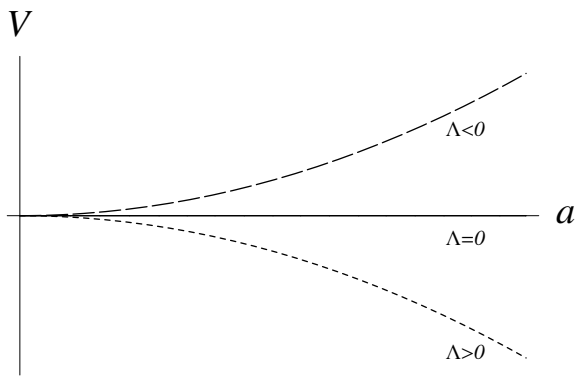}}}
\hfil }
\vskip.25cm
\caption{\small The contribution $Ba^2$ to the potential energy for different values of $\Lambda$.}
\label{F:1b}
\end{figure}

Equation~\eqref{Friedmann'} is formally equivalent~\cite{lima&co, linder} to 
\begin{equation}
\frac{\dot{a}^2}{2}+V(a)=E,
\label{totalenergy}
\end{equation}
 expressing the conservation of energy for a particle with unit mass in one dimension, with coordinate $a$ and total energy
\begin{equation}
E=-kc^2/2,
\label{E}
\end{equation}
subject to forces with potential energy
\begin{equation}
V(a)=-\frac{A}{a^{1+3w}}+B a^2,
\label{V}
\end{equation}
where 
\begin{equation}
A:=\frac{4\pi G\rho_0\,a_0^{3(1+w)}}{3 c^2}>0,\quad\quad B:=-\frac{\Lambda c^2}{6}.
\label{AB}
\end{equation}  
The two contributions to $V$ are plotted in Figs.~\ref{F:1a} and~\ref{F:1b}.  
One notes that qualitative differences arise for the critical values $w=-1/3$, $w=-2/3$, and $\Lambda=0$.  Another exceptional value is   $w=-1$, because in this case also the first contribution becomes quadratic, so $V(a)=\left(B-A\right)a^2$, with $A=4\pi G\rho_0/(3 c^2)$, and the evolution is entirely controlled by an effective cosmological constant 
\begin{equation}
\Lambda_\mathrm{e}=\Lambda+6A/c^2=\Lambda+8\pi G\rho_0/c^4.
\label{Lambdaeff}
\end{equation}

Given $V(a)$ and a value of $E$, Eq.~\eqref{totalenergy} defines a set of points in the phase plane $(a,\dot{a})$.  This set corresponds to one or more phase curves, each associated with a possible solution of the equation of motion with energy $E$.~\cite{arnold, mt}  Since the kinetic term $\dot{a}^2/2$ in Eq.~\eqref{totalenergy} is very simple, such curves can be easily drawn knowing the plot of $V(a)$.  In particular, it is worth noticing that the Friedmann equation~\eqref{Friedmann'}, hence Eq.~\eqref{totalenergy}, is invariant under time-reversal, so all the diagrams in phase space will be symmetric with respect to the horizontal axis.  As is well-known from classical mechanics, from the knowledge of the phase curves one can get some useful information about the dynamical behaviour of the system.  

There is only one place where the qualitative study applied to the cosmological case differs from the analogous discussion in particle dynamics.  Usually, for a point particle, the initial conditions affect only the value of the total energy $E$ but not the potential energy, which is fixed.  Initial conditions associated with different values of $E$ will then produce different phase curves.  For a given $V(a)$, there will be infinitely many of them.  On the contrary, in the cosmological case there are only three possible values for $E$ ($c^2/2$, $0$, or $-c^2/2$, according to the spatial geometry), so if $V(a)$ is fixed there is only a finite number of phase curves.  However, $A$ depends on $\rho_0$ and $a_0$, so it can also vary from solution to solution, which makes $V(a)$ dependent on the initial conditions.  Therefore, in order to draw a representative sample of phase curves, one may need to consider, for each value of $E$ among the three possibilities, also different values of $A$ which can lead to different qualitative behaviours.  

{From} the potential energy $V(a)$ one can also infer the behaviour of another quantity of cosmological interest: The second time derivative $\ddot{a}$ of the scale factor, which is related to the deceleration parameter $q=-a\ddot{a}/\dot{a}^2$.  Indeed, taking the time derivative of Eq.~\eqref{totalenergy} one finds 
\begin{equation}
\ddot{a}=-\frac{\mathrm{d}V}{\mathrm{d}a},
\label{Newton}
\end{equation}
which is, of course, Newton's law of motion.  With $V(a)$ given by Eqs.~\eqref{V}--\eqref{AB}, this is equivalent to Friedmann's equation in the form
\begin{equation}
\ddot{a}=-\frac{4\pi G}{3 c^2}\rho\left(1+3 w\right)a+\frac{\Lambda c^2}{3}\,a,
\label{acc}
\end{equation}
where $\rho$ is given by Eq.~\eqref{rho-a}.   By the assumption $\rho> 0$, the first contribution on the right-hand side of Eq.~\eqref{acc} is strictly negative if and only if $w>-1/3$, and vanishes for $w=-1/3$.  Hence, a positive $\ddot{a}$, as currently observed,~\cite{lambda, Linder, linder} requires a positive cosmological constant or exotic matter with $w<-1/3$.  This supports a study of FLRW models with a wide range of combinations of $w$ and $\Lambda$.


\section{Qualitative study}
\label{sec4}
\setcounter{equation}{0}

Given the existence of the critical values $w=-1/3$, $w=-2/3$, $w=-1$, and $\Lambda=0$, there are 21 possible classes of models.  We shall group them together according to the value of the parameter $w$.  However, only the physically most interesting cases will be treated in detail.  We do not discuss at length the physical meaning of the different possibilities; see Ref.~\onlinecite{nemiroff} for an instructive presentation of the physics underlying different models and for an extensive list of references.

Let us first explain the conventions used in the figures.  For each class of models considered, there are three interesting plots to show: One for the effective potential energy $V(a)$, one for the phase portrait (the set of phase curves for a given $V$), and one for the corresponding time evolutions of the scale factor $a(t)$.

The potential energy is always drawn above the corresponding phase portrait, and the units along the $a$-axis are the same for both diagrams.  We omit the arrows that are sometimes drawn on the phase curves, which point to the right when $\dot{a}>0$ (upper half of the phase plane) and to the left when $\dot{a}<0$ (lower half).  The values $k=-1$ and $k=1$ correspond to the values $c^2/2$ and $-c^2/2$ of the total energy, and to the upper and lower horizontal lines in the diagram for $V(a)$, whereas $k=0$ corresponds to $E=0$ and to the $a$-axis.  For convenience, we associate different colours with different values of $k$: Red, black, and blue for $k=-1$, $k=0$, and $k=1$, respectively.  All the curves pertaining to a value of $k$ are always drawn in the corresponding colour.  We use colours for greater clarity, but all the information can be retrieved also in grayscale.  A comparison between the phase diagram and the corresponding curve for $V(a)$ makes it very easy to understand which phase curves correspond to a given value of the energy $E$; therefore, we do not introduce any label for this purpose.  Sometimes, it is useful to draw plots for $V(a)$ with different values of the parameter $A$ (see, e.g., Fig.~\ref{F:6}).  In this case we use different styles (solid, dotted, dashed) both for such plots and for the corresponding phase curves.

The colour and style of the curves in the diagrams for $a(t)$ are the same as for the corresponding phase curves.  Since here the correspondence with the value of $E$ is less obvious in greyscale, we have added a few labels.  Whenever all the three curves $a(t)$ corresponding to the three possibilities $k=-1$, $k=0$, and $k=1$ are drawn for the same $V(a)$, they are identified by a ``$-$'', no label, and a ``$+$'', respectively.  From the diagrams below one can see that this labelling involves only solid curves.  Whenever there are two disconnected curves that are related by time-reversal, corresponding to two alternative histories of the universe with opposite signs for $\dot{a}$ (as it happens, for example, for $w>-1/3$ and $\Lambda=0$ in the cases $k=-1$ and $k=0$, which admit solutions that contract to a big crunch together with those that expand from a big bang, see Sec.~\ref{Sss:separatrix} below), only the one with $\dot{a}>0$ will be plotted.  The other (physically irrelevant) curve $a(t)$ with $\dot{a}<0$ can be obtained simply by reflection through a vertical axis.  Also, note that the time $t=0$, as well as the value of $t$ corresponding to the big bang, has nothing special, because the Friedmann equation~\eqref{Friedmann'} and Eq.~\eqref{totalenergy} are invariant under time translations.  This allows us to plot the curves starting from the same point (as in Fig.~\ref{F:3}), but when this is done, it is only for graphical convenience.  For the same reason, possible intersections between curves have no physical meaning.

\subsection{$w>-1/3$}
\label{Ss:13}

This applies to all cases of ``ordinary'' matter, including in particular dust ($w=0$) and radiation ($w=1/3$), but also to several exotic substances, like e.g.~stiff matter ($w=1$).  Since $1+3w>0$, the potential energy $-A/a^{1+3w}$ is singular at $a=0$, increases monotonically with $a$, and tends to 0 for $a\to +\infty$.  In all these models there is a big bang and/or a big crunch, i.e., $a=0$ at some time $t^-$ in the past and/or $t^+$ in the future.  For $a\to 0$, the terms in $\Lambda$ and in $k$ are negligible in Eq.~\eqref{Friedmann'}, which becomes asymptotically
\begin{equation}
\frac{\dot{a}^2}{2}\sim\frac{4\pi G\rho_0\,a_0^{3(1+w)}}{3 c^2}\,\frac{1}{a^{1+3w}}\;.
\label{Friedmann-a0}
\end{equation}
This is immediately integrated to obtain
\begin{equation}
a(t)\sim a_0\left(\frac{6\pi G\rho_0(1+w)^2}{c^2}\right)^{\frac{1}{3(1+w)}} 
\left(\pm\left(t^\pm-t\right)\right)^{\frac{2}{3(1+w)}}\!,
\label{abb}
\end{equation}
valid for $t$ close to $t^\pm$.  The negative signs corresponds to the big bang, the positive signs to the big crunch.  As the big bang/big crunch are approached, $\dot{a}\to \pm\infty$.

\subsubsection{$\Lambda=0$}
\label{Sss:separatrix}

\begin{figure}
\vbox{\hfil
\scalebox{0.65}{\includegraphics{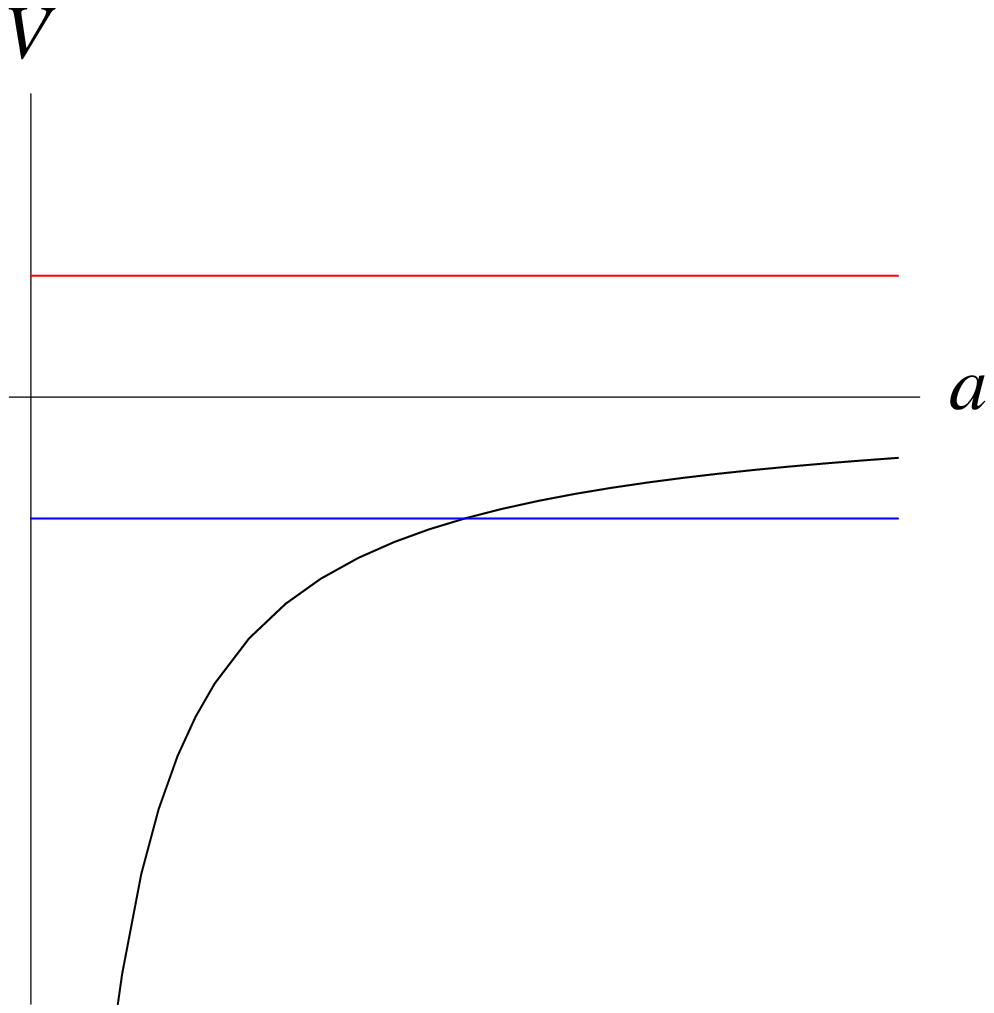}}  \vskip.5cm \scalebox{0.65}{\hskip1cm \includegraphics{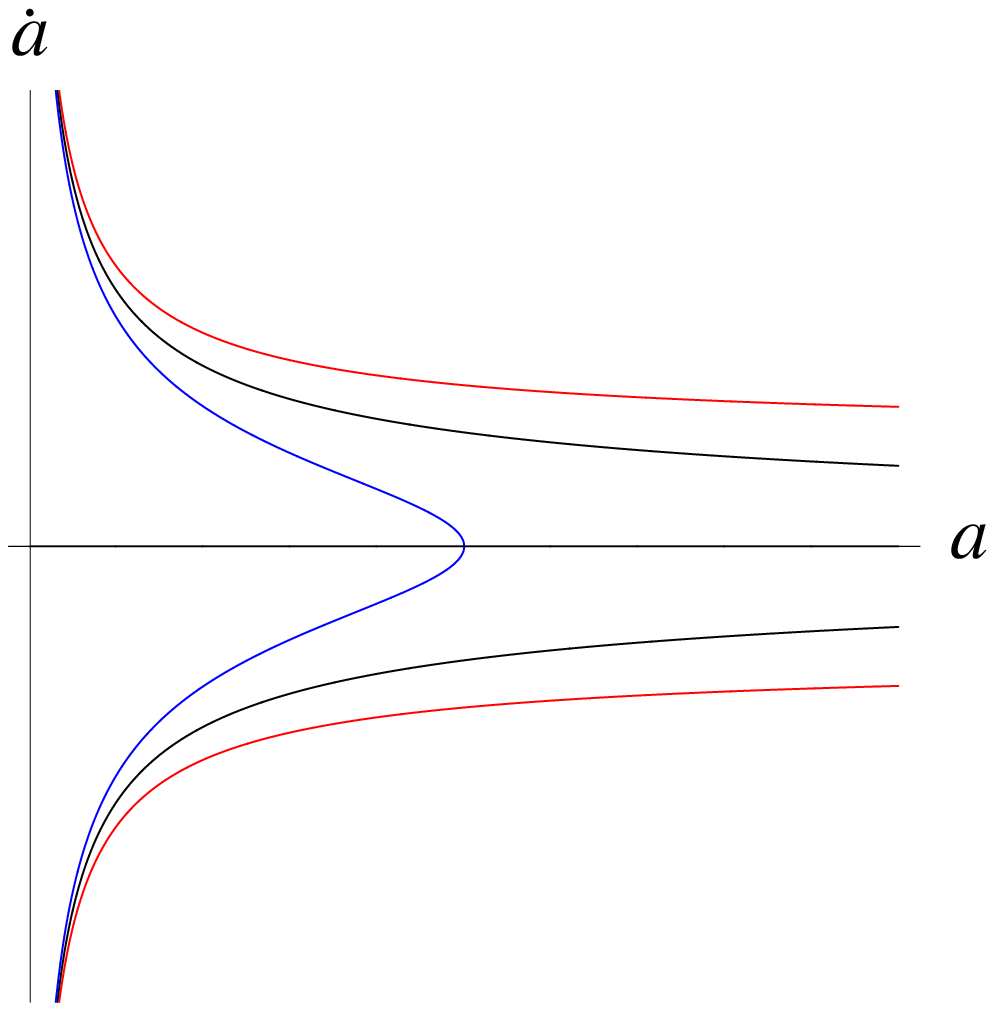}}
\hfil}
\vskip.25cm
\caption{\small Potential energy and phase portrait for $w>-1/3$, $\Lambda=0$.}
\label{F:2}
\end{figure}

\begin{figure}[tbp]
\vbox{ \hfil
\scalebox{0.8}{\includegraphics{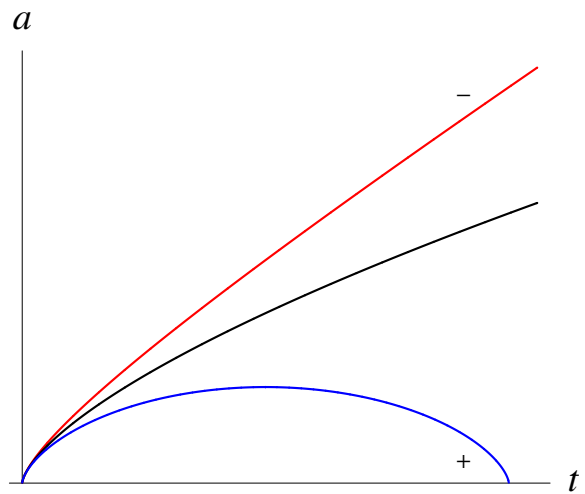}}
\hfil }
\vskip.25cm
\caption{\small Time dependence of the scale factor corresponding to the phase curves of Fig.~\ref{F:2}.}
\label{F:3}
\end{figure}

The potential energy and the phase portrait are shown in Fig.~\ref{F:2}.
For $k=1$ the scale factor $a(t)$ starts from zero with a big bang, reaches a maximum, and then symmetrically recollapses to zero in a big crunch.  For $k=0$ or $k=-1$, the expansion continues forever but slows down, and for $a\to +\infty$, $\dot{a}\sim -kc$.  There is also a physically irrelevant time-reversed process in which $\dot{a}<0$.  Interestingly, the diagram for the three possible behaviours of $a(t)$, see Fig.~\ref{F:3}, is qualitatively indistinguishable from those presented in many textbooks for the case $w=0$, although here the value of $w>-1/3$ is arbitrary.

The curves corresponding to $k=0$ in the phase portrait of Fig.~\ref{F:2} and in Fig.~\ref{F:3} mark the boundary between two qualitatively different regimes --- one in which models recollapse and one in which models expand forever.  In general, qualitatively different behaviours occur when the potential energy has maxima (even improperly, when it is attained for $a\to +\infty$, as in Fig.~\ref{F:2}).  We shall call separatrix the set of points in the phase space corresponding to a total energy equal to such a maximum value of $V$.  This set always separates regions of the phase space containing phase curves corresponding to qualitatively different behaviours.

\subsubsection{$\Lambda<0$}

\begin{figure}[htbp]
\vbox{\hfil
\scalebox{0.65}{\includegraphics{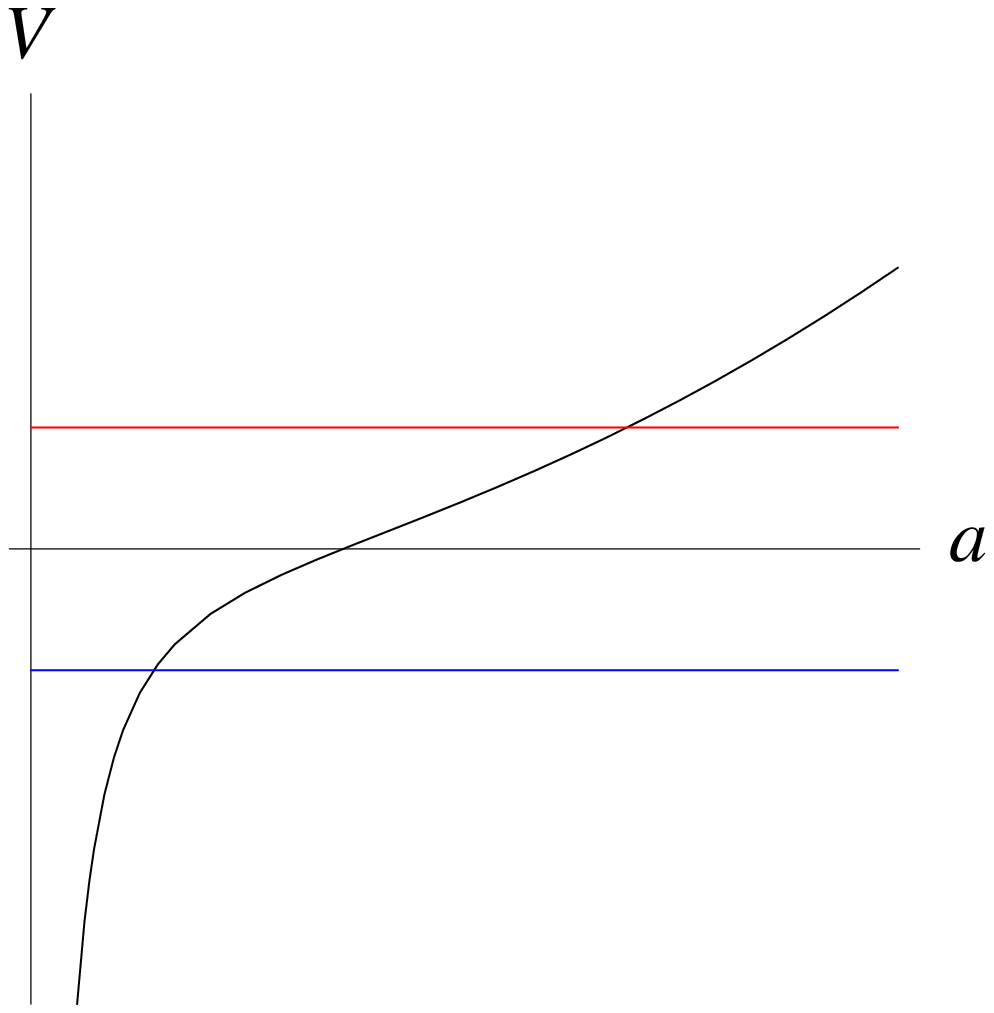}}  \vskip.5cm \scalebox{0.65}{\hskip1cm \includegraphics{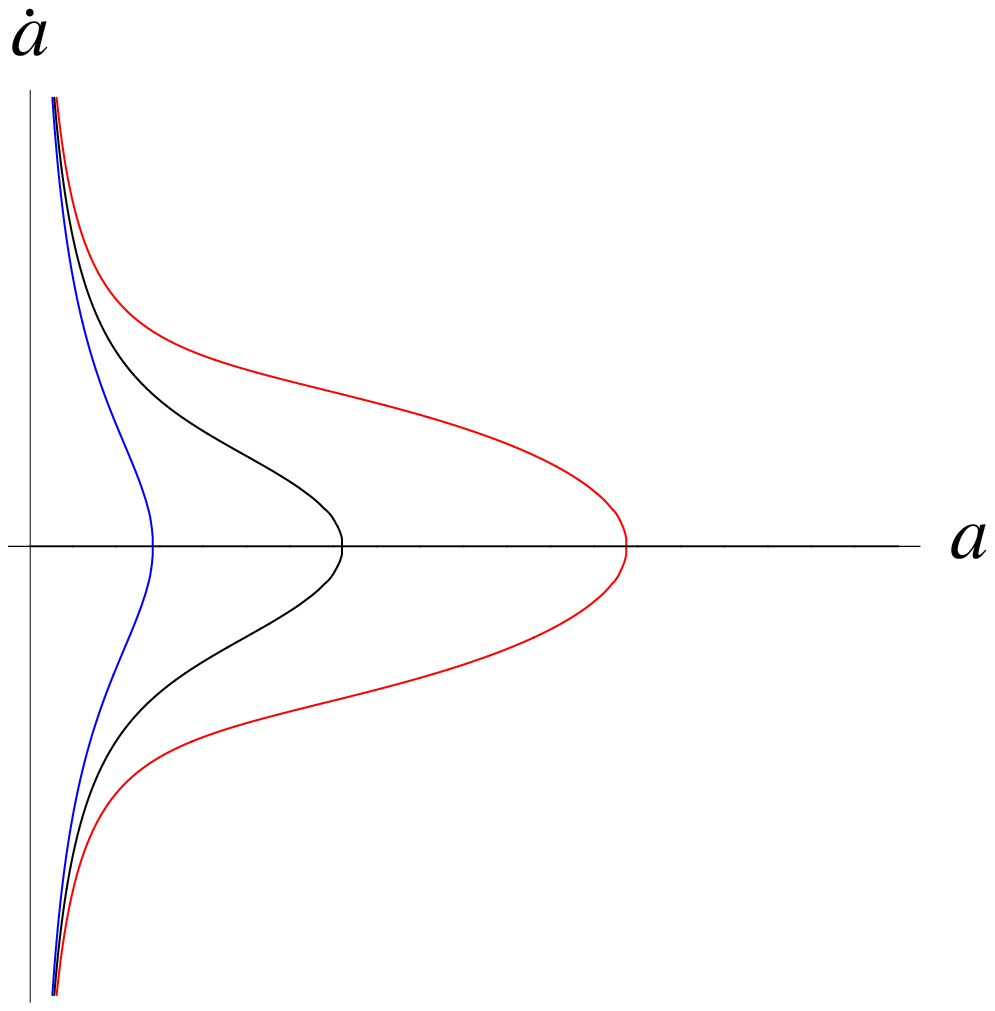}}
\hfil}
\vskip.25cm
\caption{\small Potential energy and phase portrait for $w>-1/3$, $\Lambda<0$.  In all cases, the time dependence of the scale parameter is similar to the one labeled by a ``+'' in Fig.~\ref{F:3}.}
\label{F:4}
\end{figure}

A negative cosmological constant corresponds, in Newtonian language, to an attractive universal force.  The potential energy is the one plotted in the upper part of Fig.~\ref{F:4}, and leads to the phase portrait shown in the lower part.
The phase of expansion is always followed by a recollapse and a big crunch (so the plots of $a(t)$ all look, qualitatively, like the one in Fig.~\ref{F:3} that is labeled by ``$+$''), independently of the value of $k$.  The asymptotic behaviour near the big bang and the big crunch is still given by Eq.~\eqref{abb}.

\subsubsection{$\Lambda>0$}
\label{S:Einstein}

\begin{figure}[htbp]
\vbox{\hfil
\scalebox{0.65}{\includegraphics{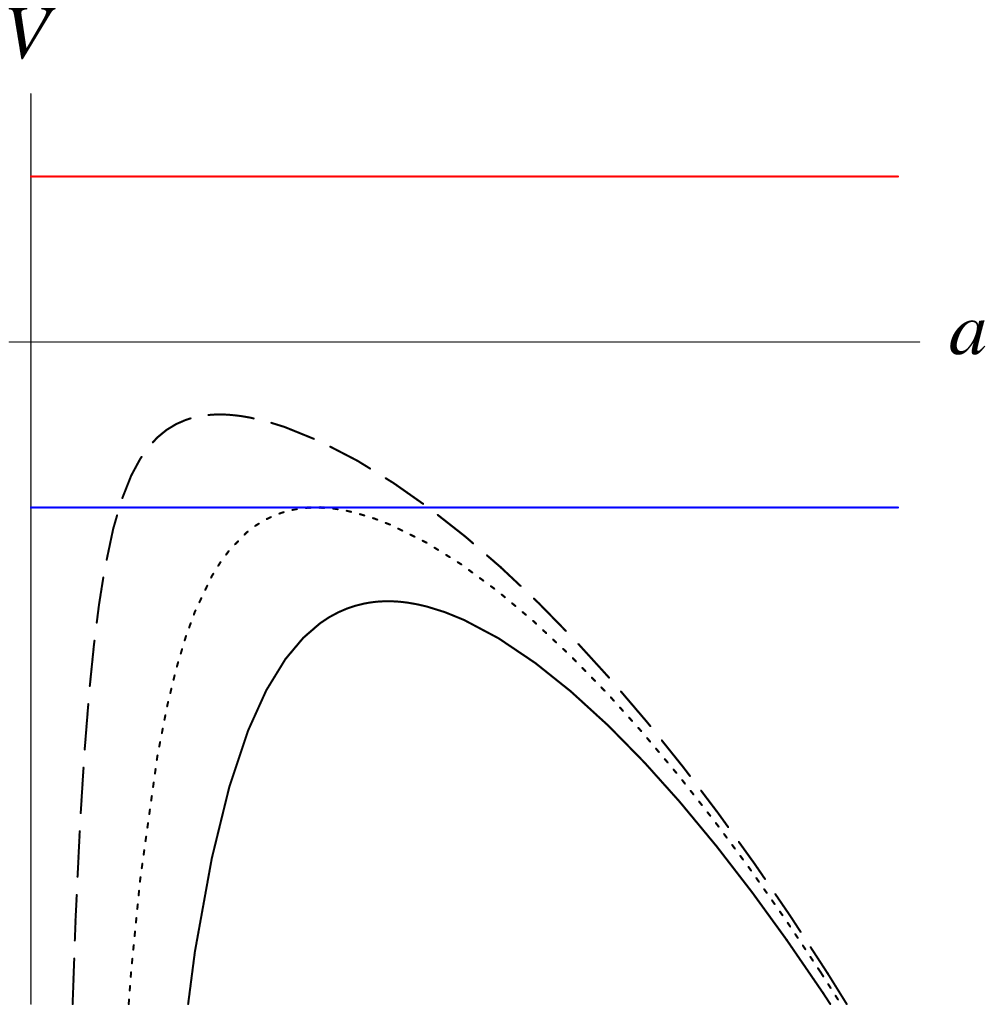}}  \vskip.5cm \scalebox{0.65}{\hskip1cm \includegraphics{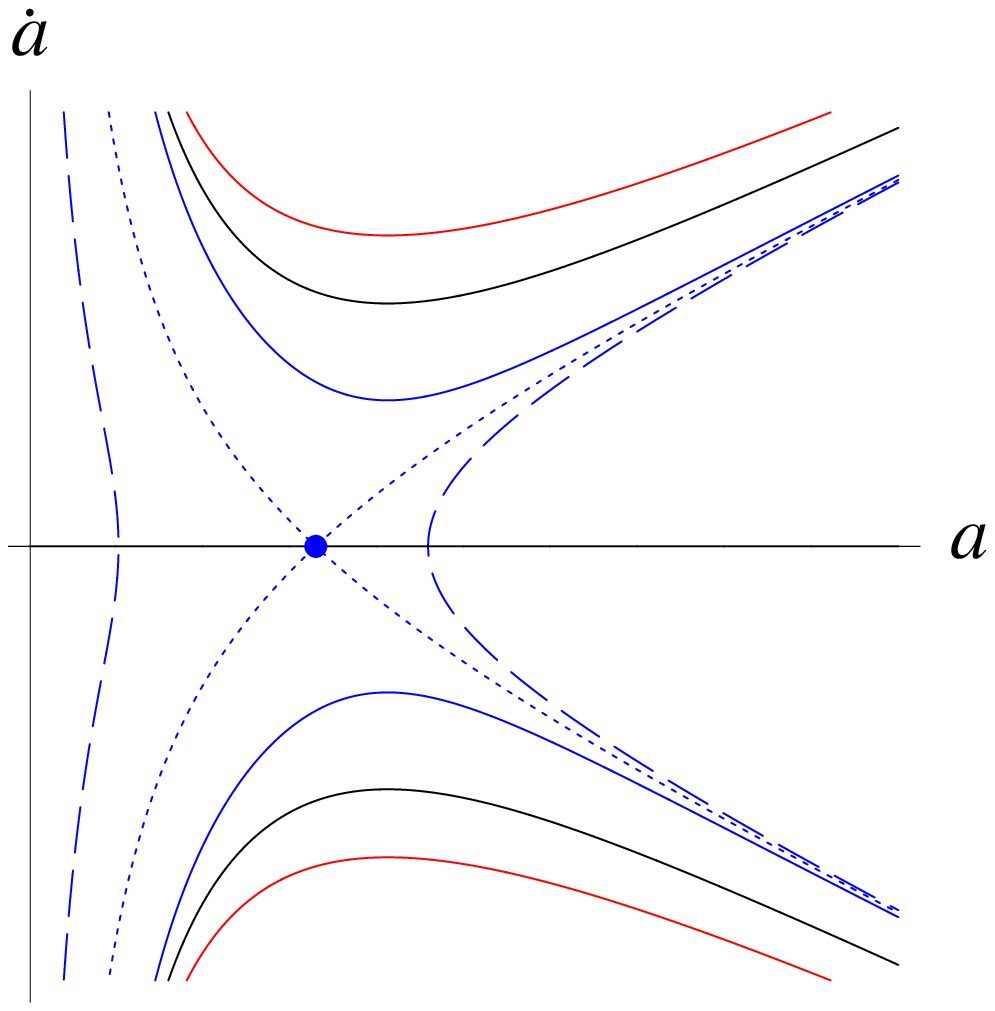}}
\hfil}
\vskip.25cm
\caption{\small Potential energy and phase portrait for $w>-1/3$, $\Lambda>0$ , and three possible values of $A$.  The phase curves corresponding to $k=-1$ and $k=0$ are drawn only for the potential energy represented by the solid curve, since in the other two cases they look qualitatively the same.}
\label{F:6}
\end{figure}

\begin{figure}[htbp]
\vbox{ \hfil
\scalebox{0.8}{\includegraphics{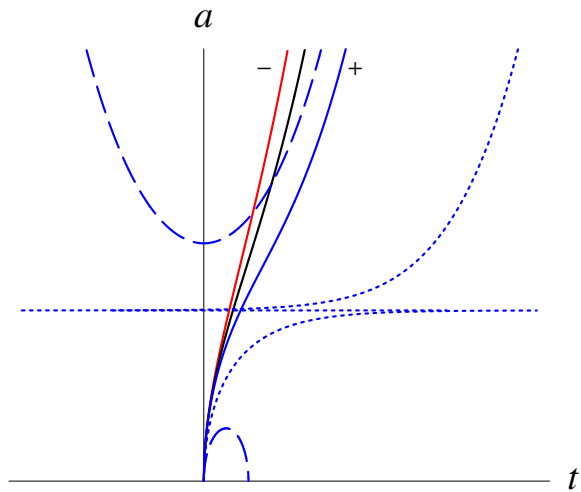}}
\hfil }
\vskip.25cm
\caption{\small Time dependence of the scale factor corresponding to the phase curves of Fig.~\ref{F:6}.}
\label{F:8}
\end{figure}

The potential energy is plotted in the upper part of Fig.~\ref{F:6}.  
It is always negative, and for $a=\bar{a}$, with 
\begin{eqnarray}
\lefteqn{\mbox{\hskip-5truecm} \bar{a}=\left(\frac{(1+3w)A}{-2B}\right)^{\frac{1}{3(1+w)}}}\nonumber\\
=a_0\left(\frac{4\pi (1+3w)G\rho_0}{\Lambda c^4}\right)^{\frac{1}{3(1+w)}},
\label{bara}
\end{eqnarray}
attains its maximum value
\begin{eqnarray}
\lefteqn{\mbox{\hskip-4.15truecm} V(\bar{a})=-\frac{3(1+w)}{2}\,A^{\frac{2}{3(1+w)}}\left(\frac{-2B}{1+3w}\right)^{\frac{1+3w}{3(1+w)}}}\nonumber\\
\mbox{\hskip-.5truecm}=-\frac{1+w}{2(1+3w)}\,\bar{a}^2\Lambda c^2.
\label{Vbar}
\end{eqnarray}
The phase portrait is shown in the lower part of Fig.~\ref{F:6}.  For $k=-1$ and $k=0$ the universe starts with a big bang, with asymptotic behaviour given by~\eqref{abb}, and traverses a phase in which $\dot{a}$ decreases, followed by one in which it increases.  Asymptotically, for large values of $a$, Eq.~\eqref{Friedmann'} becomes
\begin{equation}
\frac{\dot{a}^2}{2}\sim \frac{\Lambda c^2}{6}\,a^2
\end{equation}
and gives the de Sitter expansion 
\begin{equation}
a(t)\sim K\mathrm{e}^{\sqrt{\Lambda/3}\;\, ct},
\label{deSitter}
\end{equation}
where $K$ is a constant (see the two leftmost solid curves in Fig.~\ref{F:8}).
Of course, there are also the time-reversed solutions (not shown in Fig.~\ref{F:8}) in which the universe contracts from an asymptotic de Sitter phase at $t\to -\infty$ and ends with a big crunch.

For $k=1$ the situation is more complex, because the behaviour of the scale factor depends on the specific values of the parameters.  There are three qualitatively different possibilities, according to whether the maximum value $V(\bar{a})$ of the potential energy is greater, equal, or smaller than $-c^2/2$.  

If $V(\bar{a})>-c^2/2$ (which is, of course, an algebraic condition involving $a_0$, $\rho_0$, and $\Lambda$, and corresponds to the dashed curves in Fig.~\ref{F:6}), there are two types of behaviour, according to the initial conditions.  Either the universe starts with a big bang, reaches a maximum size (smaller than $\bar{a}$), and then recollapses in a big crunch, as in the models with $\Lambda\geqslant 0$, or it starts at $t\to -\infty$ with an asymptotic de Sitter phase in which $a$ decreases as 
\begin{equation}
a(t)\sim K\mathrm{e}^{-\sqrt{\Lambda/3}\;\, ct},
\label{deSitter-}
\end{equation}
reaches a state where $a$ has a minimum value (greater than $\bar{a}$), and then bounces towards another asymptotic de Sitter phase for $t\to +\infty$, in which $a(t)$ increases according to~\eqref{deSitter}.  In Fig.~\ref{F:8} these time dependences are represented by the two dashed curves.

For $V(\bar{a})=-c^2/2$ (dotted curves in Fig.~\ref{F:6}) there is a separatrix in the phase plane.  This is the union of five phase curves, one of which is degenerate (i.e., reduces to a single point; see the dot in Fig.~\ref{F:6}) and corresponds to an unstable  static model with $a(t)\equiv \bar{a}=a_0$.  From Eq.~\eqref{bara}, this happens when 
\begin{equation}
\rho_0=\frac{\Lambda c^4}{4\pi (1+3w) G}\;,
\end{equation}
whereas Eq.~\eqref{Vbar}, together with the condition $V(\bar{a})=-c^2/2$, gives 
\begin{equation}
a_0=\left(\frac{1+3w}{(1+w) \Lambda}\right)^{1/2}\;.
\label{Einsteinstatic}
\end{equation}
Hence, the value of $\Lambda$ determines uniquely both the scale factor and the energy density in this model.  For $w=0$ one recovers Einstein's static universe.~\cite{Einstein}  The other phase curves pertaining to this case correspond to models that start with a big bang, or with an asymptotic collapsing de Sitter phase, and end up approaching asymptotically the previous static universe for $t\to +\infty$, or with their time-reversed version.  In particular, the case in which a model that is static in the asymptotic past  evolves towards an expanding de Sitter universe was discovered by Lema\^{\i}tre,~\cite{lemaitre} and supported by Eddington.~\cite{eddington, eddington2}  (See also Ref.~\onlinecite{stability} for a modern assessment of the stability of the static model.)  In Fig.~\ref{F:8} these time dependences of the scale factor are represented by the dotted curves.  As already announced, only the curves with $\dot{a}\geqslant 0$ are shown.

\begin{figure}[htbp]
\vbox{ \hfil
\scalebox{0.8}{\includegraphics{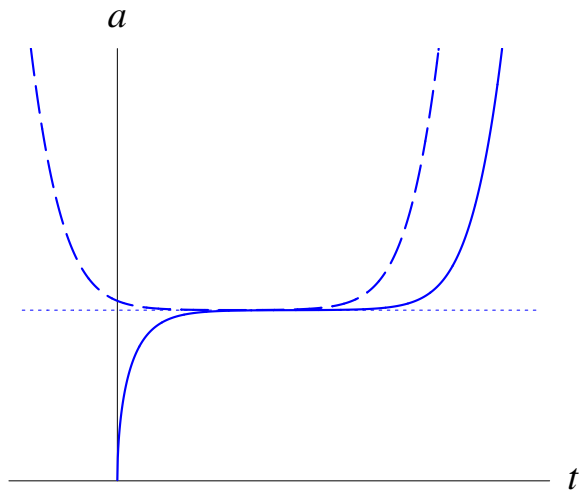}}
\hfil }
\vskip.25cm
\caption{\small Time dependence of the scale factor in coasting models.  Solid curve: $V(\bar{a})\lesssim -c^2/2$ (Lema\^{\i}tre universe).  Dashed line: $V(\bar{a})\gtrsim -c^2/2$.}
\label{F:Lemaitre}
\end{figure}

Finally, if $V(\bar{a})<-c^2/2$ (solid curves in Fig.~\ref{F:6}, and curve labelled by a ``$+$'' in Fig.~\ref{F:8}), we have a situation qualitatively similar to the one discussed in the cases $k=-1$ and $k=0$.  The only noteworthy difference is that now it is possible for $V(\bar{a})$ to be arbitrarily close to $-c^2/2$.  Hence, there are models that start with a big bang and end with an asymptotic de Sitter expansion, but have an intermediate phase where $\dot{a}$ is very close to zero, so $a(t)$ remains close to $\bar{a}$ for a very long time (solid line in Fig.~\ref{F:Lemaitre}) and there is a phase in which the universe ``coasts'' Einstein's  static model.  
This possibility was also discovered by Lema\^{\i}tre;~\cite{lemaitre2} see Ref.~\onlinecite{felten} for a detailed discussion.  Of course, other models exhibiting such a coasting can be found, all corresponding to phase curves close to the separatrix (see, e.g., the dashed line in Fig.~\ref{F:Lemaitre}).  

In the case $k=0$, it is also possible for the maximum value of $V$ to be very close to $0$ (i.e., to $E$).  However, the only possibility for a static solution in this case is $\bar{a}=0$ (which happens when the ratio $A/B$ attains the degenerate value $A/B=0$).  Therefore, a quasi-static phase in which $a$ is almost constant, like in Lema\^{\i}tre coasting models, is possible only right after the big bang, when the universe is extremely dense.  At the end of such a phase, the universe turns eventually into a de Sitter model.

\subsection{$-1\leqslant w\leqslant -1/3$}
\label{Ss:strings}

The cases $-1/3<w<-2/3$, $w=-2/3$, and $-2/3<w<-1$, although mathematically distinct, exhibit many similarities.  They all correspond to a fluid of topological defects (strings or domain walls).~\cite{nemiroff}  We do not discuss all the different possibilities, but present only the case with $-1<w<-2/3$ and $\Lambda<0$, which exhibits interesting  behaviours.  It is an easy exercise to draw the phase curves and discover the behaviour of $a(t)$ for other values of the parameters.  

\begin{figure}[htbp]
\vbox{\hfil
\scalebox{0.65}{\includegraphics{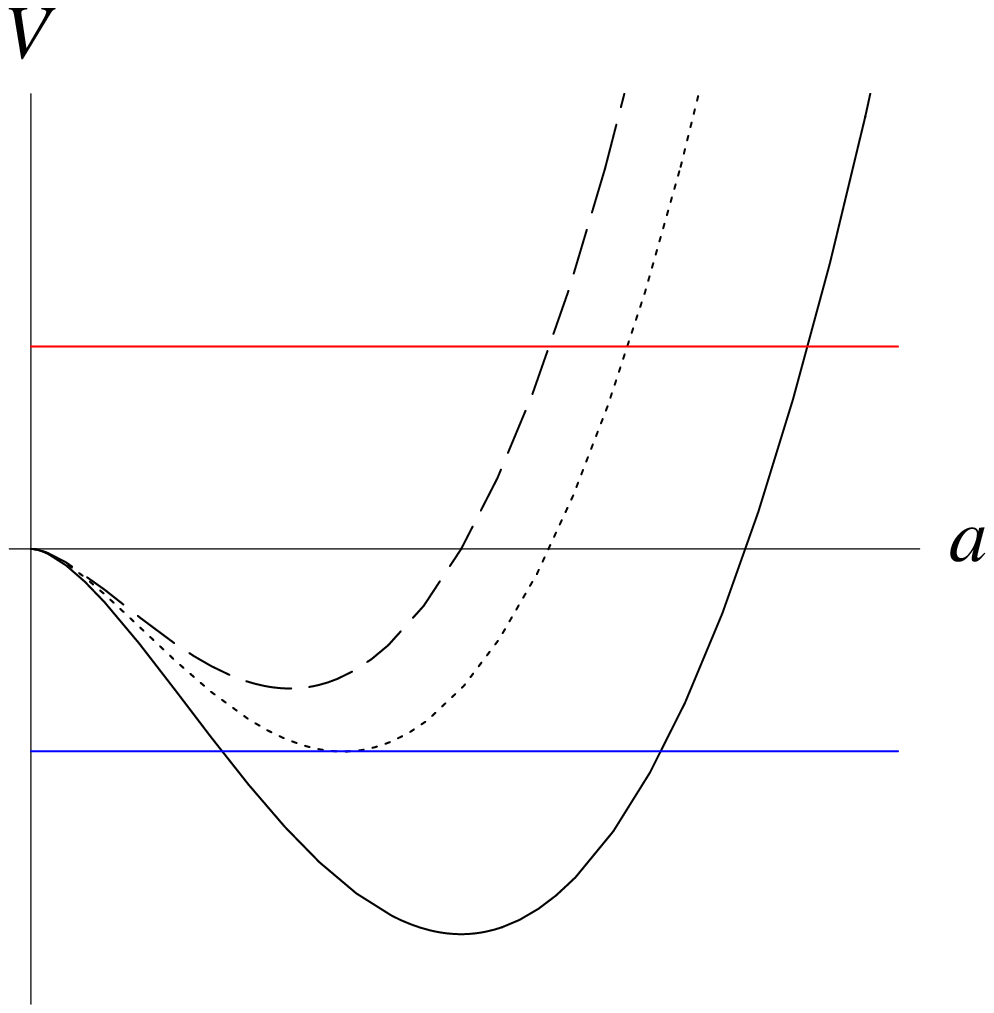}}
\hskip.5cm \scalebox{0.65}{\hskip1cm
\includegraphics{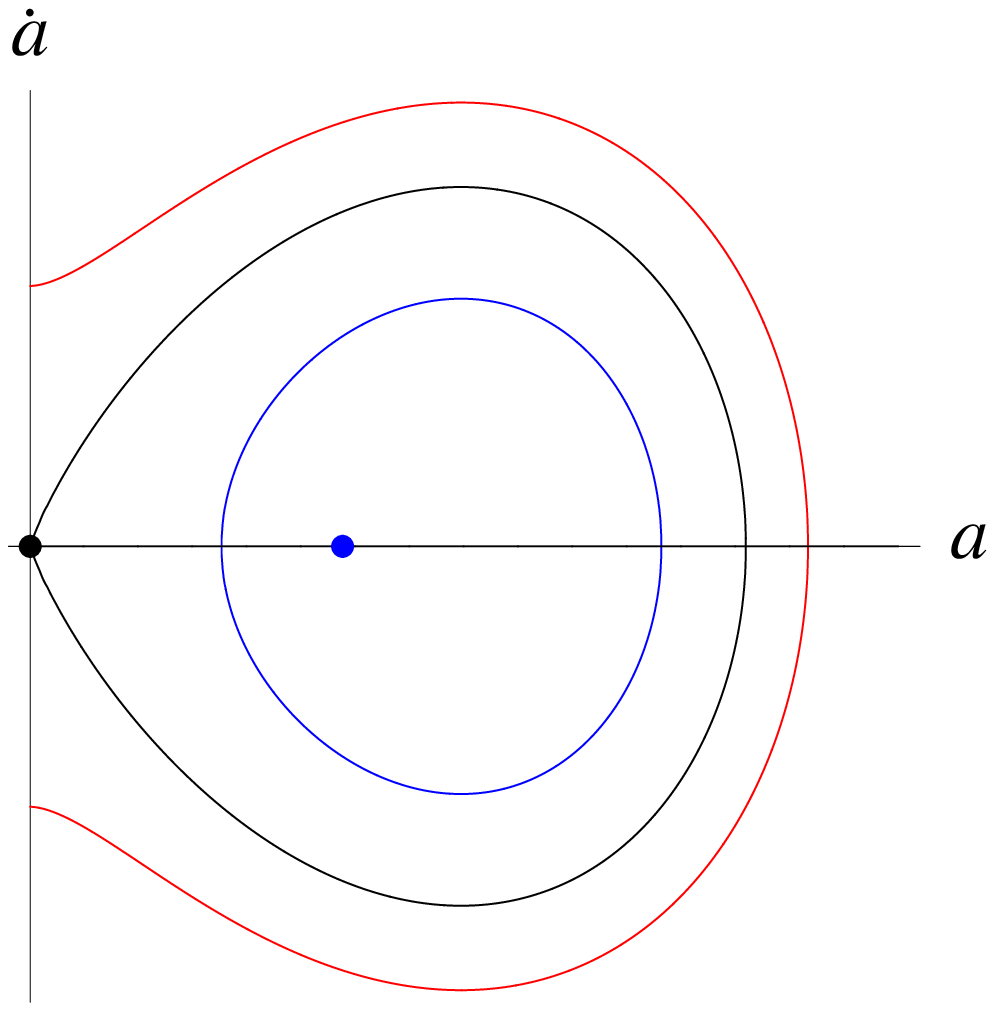}} \hfil} \vskip.25cm
\caption{\small Potential energy and phase portrait for $-1<w<-2/3$, $\Lambda<0$, and three values of $A$.  With the exception of the dot at $a\neq 0$, which denotes the stable static model when $V$ is represented by the dotted curve, only the phase curves for the largest value of $A$ have been  drawn (solid curve $V(a)$), because the same qualitative behaviour will occur in the other cases.}
 \label{F:perepe}
\end{figure}

\begin{figure}[htbp]
\vbox{ \hfil
\scalebox{0.8}{\includegraphics{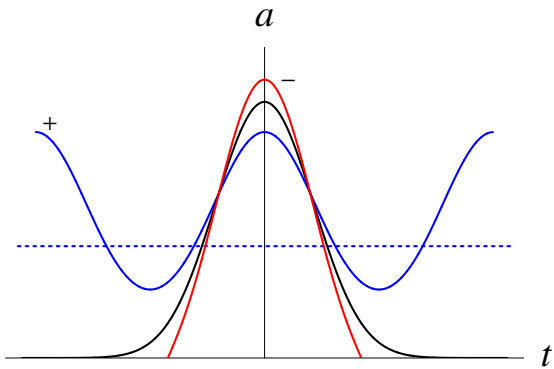}} \hfil }
\vskip.25cm \caption{\small Time dependence of the scale factor
corresponding to the phase curves of Fig.~\ref{F:perepe}.} \label{F:param-perepe}
\end{figure}

The potential energy and the corresponding phase curves are plotted in Fig.~\ref{F:perepe}.  Figure~\ref{F:param-perepe} shows the time behaviour of $a$.  For $k=-1$, there is always a big bang followed by a big crunch.  Note that, contrary to what happened when $w>-1/3$, now $\dot{a}$ is finite at these events.  For $k=0$ (separatrix) there is either a physically uninteresting singular ``universe'' with $a(t)\equiv 0$ (dot at the origin of the phase plane), or a solution where the big bang and big crunch take place infinitely far in time.  The behaviour in the case $k=1$ depends on the minimum value of the potential energy, attained at $\bar{a}$ given by Eq.~\eqref{bara}.  Such minimum value is, of course, still given by Eq.~\eqref{Vbar}.  If $V(\bar{a})>-c^2/2$ (dashed curve for the potential energy in Fig.~\ref{F:perepe}), no solution exists.  If $V(\bar{a})=-c^2/2$, there is a stable static model with scale factor given by Eq.~\eqref{Einsteinstatic}~\cite{fn5} (dotted curves for $V(a)$ and $a(t)$, and dot with $a\neq 0$ in the phase portrait).  Finally, if $V(\bar{a})<-c^2/2$ (solid curve for the potential energy) the universe oscillates around the static model.


Let us now consider the border cases $w=-1$ and $w=-1/3$.  For a fluid with $w=-1$ the stress-energy-momentum tensor~\cite{newfn}
\begin{equation}
T_{ab}=\left(\rho+p\right)u_a u_b+p\,\mathrm{\sl g}_{ab}
\end{equation}
reduces to $T_{ab}=-\rho\,\mathrm{\sl g}_{ab}$.  This circumstance is interesting for two reasons.  First of all, the four-velocity of the fluid does not appear in $T_{ab}$, so the observable properties of the fluid do not allow one to identify a preferred reference frame.  In particular, this is the case for the quantum vacuum,~\cite{carroll, earman, lambda, vacuum} which must, of course, be Lorentz-invariant.  Second, the condition $\nabla^b T_{ab}=0$, expressing the local conservation of energy and momentum, implies that $\rho$ is a constant.~\cite{conservation}  Inserting the stress-energy-momentum tensor $T_{ab}=-\rho\,\mathrm{\sl g}_{ab}$ into Einstein's equations, one then sees that such a fluid behaves like a cosmological constant with value $\Lambda_\ast=8\pi G\rho_0/c^4$.  The evolution of the universe is then entirely controlled by the effective cosmological constant $\Lambda_\mathrm{e}=\Lambda+\Lambda_\ast$ as in Eq.~\eqref{Lambdaeff}, and the potential energy becomes 
\begin{equation}
V(a)=-\frac{\Lambda_\mathrm{e}\, c^2}{6}\,a^2.
\end{equation}
Similarly, for $w=-1/3$, corresponding to a universe dominated by a network of cosmic strings,~\cite{nemiroff} the potential energy is  
\begin{equation}
V(a)=-\frac{\Lambda\, c^2}{6}\,a^2-A.
\end{equation}
We do not discuss the behaviours corresponding to these simple forms of $V(a)$, which are well-known from elementary mechanics.

\subsection{$w<-1$}
\label{Ss:phantom}

This is the ``phantom energy'' regime,~\cite{nemiroff, caldwell, bigrip} introduced as an explanation for the observed acceleration of the universe.~\cite{Linder, linder}  Now $1+3w<-2$ so, defining $\delta=-3(1+w)/2>0$ such that $(1+3w)=-2(1+\delta)$, the first contribution to potential energy is 
\begin{equation}
-Aa^{2(1+\delta)}.
\end{equation}
Regardless of the value of $\Lambda$, this term dominates the potential energy for large values of $a$.  In that regime, the total energy $E$ is also negligible, and one can write the asymptotic differential equation 
\begin{equation}
\frac{\dot{a}^2}{2}\sim Aa^{2(1+\delta)}, 
\end{equation}
whose solution is 
\begin{equation}
a(t)\sim \left(\pm\delta\,\sqrt{2A}\left(t^\pm_\mathrm{br}-t\right)\right)^{-1/\delta},
\label{bigrip}
\end{equation}
where $t_\mathrm{br}^\pm$ are constants of integration.  Hence, at some finite time in the future (positive signs in Eq.~\ref{bigrip}) or in the past (negative signs), the scale factor becomes infinite.  The former event is usually referred to as the ``big rip''.~\cite{bigrip} 

It is interesting to estimate $t^+_\mathrm{br}$.  If $t_0$ is a time at which phantom energy dominates over all other forms of matter (including, possibly, a cosmological constant), equation~\eqref{bigrip} with the positive signs is correct with excellent approximation at time $t_0$, so 
\begin{equation}
a_0\approx \left(\delta\,\sqrt{2A}\left(t^+_\mathrm{br}-t_0\right)\right)^{-1/\delta}.
\label{ast}
\end{equation}
Hence, 
\begin{equation}
t^+_\mathrm{br}-t_0\approx \frac{1}{\delta\, a_0^\delta\,\sqrt{2A}}=\frac{1}{\delta}\left(\frac{8\pi G \rho_0}{3 c^2}\right)^{-1/2},
\label{delta}
\end{equation}
where we have substituted $A$ from the first equation in~\eqref{AB}.  If the cosmological constant and the spatial curvature (proportional to the numerical parameter $k$) are negligible at time $t_0$, Eqs.~\eqref{Friedmann} and~\eqref{delta} imply that
\begin{equation}
t_\mathrm{br}^+-t_0\approx\frac{1}{\delta}\,\frac{a_0}{\dot{a}(t_0)}.
\end{equation}
Hence, if $w$ is not too close to $-1$ (so $\delta$ is not too close to zero), $t_\mathrm{br}^+-t_0$ is of the same order of magnitude of the quantity $a_0/\dot{a}(t_0)$.  This is the inverse of the Hubble parameter, which is comparable, in the standard model of cosmology, with the time elapsed since the big bang.~\cite{liddle}  Choosing $t_0$ of the order of the present time (at which the cosmic acceleration starts to be relevant), it then turns out that $t_\mathrm{br}^+-t_0$ is of the same order of magnitude of the present age of the universe (or smaller, if $w$ has a large negative value).  Therefore, the big rip happens in a not-too far future (in reference~\onlinecite{bigrip} an estimate of about $20$ billion years for $t^+_\mathrm{br}-t_0$ is given assuming $w=-3/2$).  On the other hand, for $w$ very close to $-1$, $\delta$ is very close to zero, and the big rip is delayed to a remote future.~\cite{littlerip}

It may seem pointless to consider combinations of $\Lambda$ and phantom energy, which in cosmological models of current interest  both serve to explain cosmic acceleration at the present epoch.  However, there are good independent reasons to believe that $\Lambda\neq 0$,~\cite{carroll, earman, lambda} and if one admits the possibility of a component with $w<-1$ one can envisage models in which at earlier times $\Lambda$ dominates over phantom energy, which later takes over and controls the time evolution in the remote future.  For this reason, we present all such cases, again with the obvious caveat that none of them applies to the evolution of the universe in its entirety.

\subsubsection{$\Lambda\geqslant 0$}

\begin{figure}[htbp]
\vbox{\hfil
\scalebox{0.65}{\includegraphics{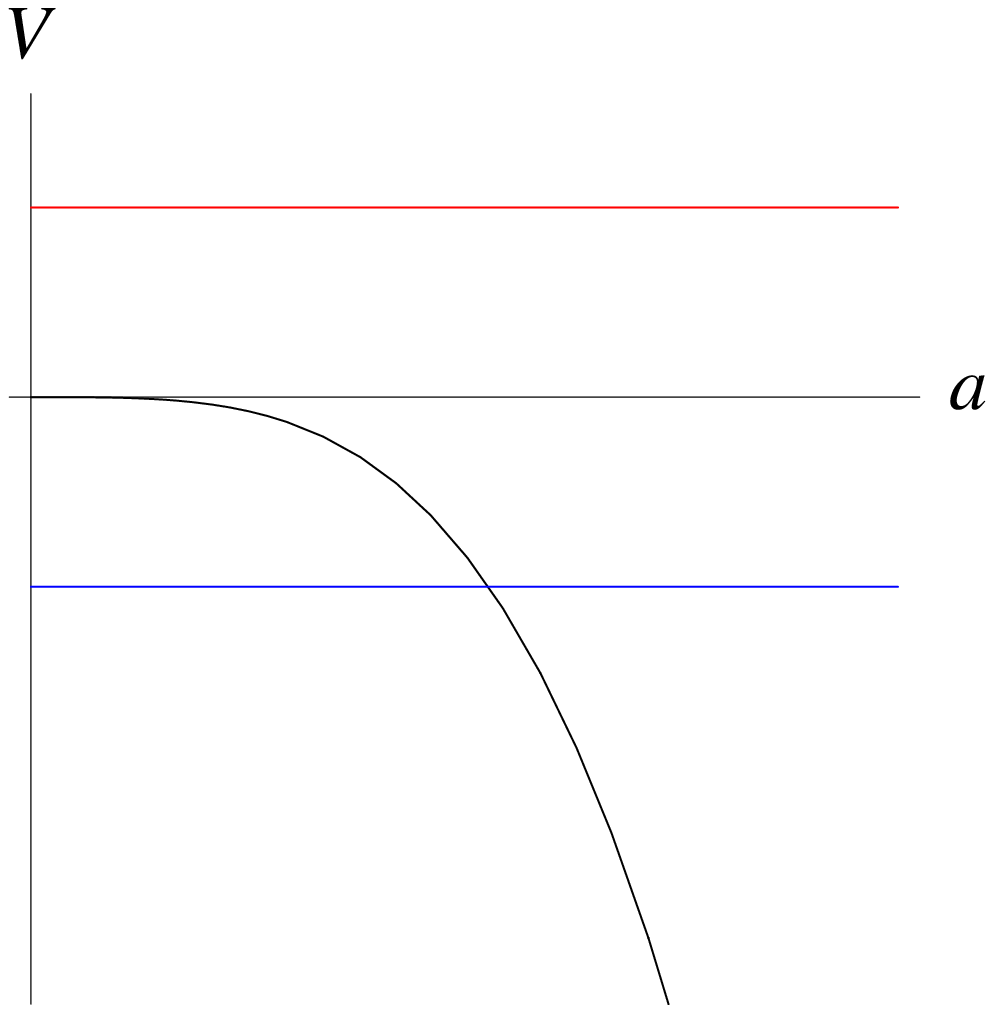}}  \vskip.5cm \scalebox{0.65}{\hskip1cm \includegraphics{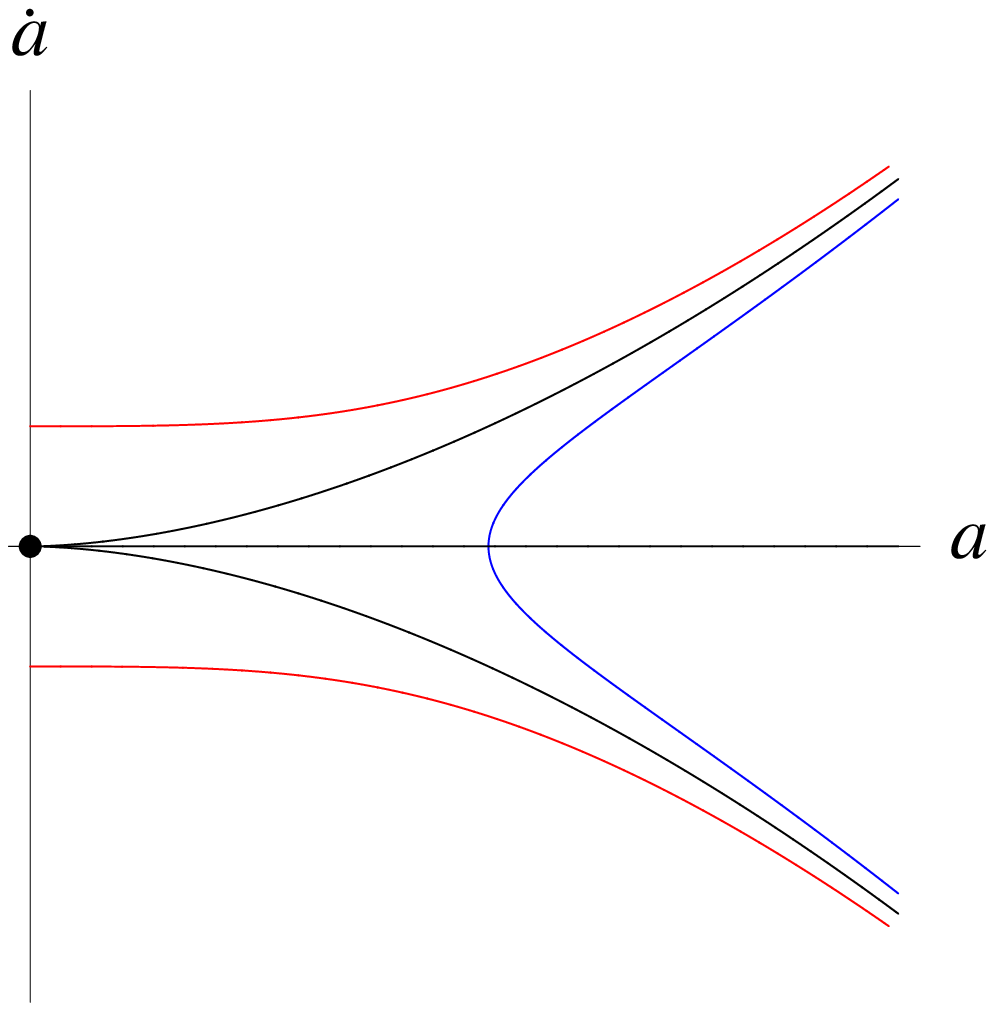}}
\hfil}
\vskip.25cm
\caption{\small Potential energy and phase portrait for $w<-1$, $\Lambda\geqslant 0$.}
\label{F:lulu}
\end{figure}

\begin{figure}[tbp]
\vbox{ \hfil
\scalebox{0.8}{\includegraphics{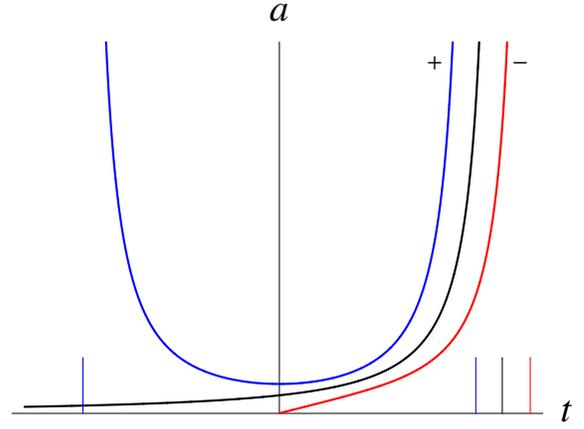}}
\hfil }
\vskip.25cm
\caption{\small Time dependence of the scale factor corresponding to the phase curves of Fig.~\ref{F:lulu}.  Note the presence of vertical asymptotes (corresponding to the ticks on the $t$-axis) associated with a big rip and its time-reversed version.}
\label{F:param-bigrip}
\end{figure}

The potential energy and the phase portrait are shown in Fig.~\ref{F:lulu}, 
and the time evolution of the scale factor is plotted in Fig.~\ref{F:param-bigrip}.   
For $k=-1$ the universe starts with a big bang and ends with a big rip, while for $k=0$ (separatrix) the big bang happens infinitely far in the past (of course, mathematically, also the time-reversed processes are possible).  For $k=0$ there is also the unphysical possibility of a static singular ``universe'' with $a(t)\equiv 0$, represented by the dot at the origin of the phase plane.  For $k=1$ the universe starts with an ``anti-big rip'', contracts to a minimum size in a finite time and then re-expands symmetrically.

\subsubsection{$\Lambda<0$}

\begin{figure}[tbp]
\vbox{\hfil
\scalebox{0.65}{\includegraphics{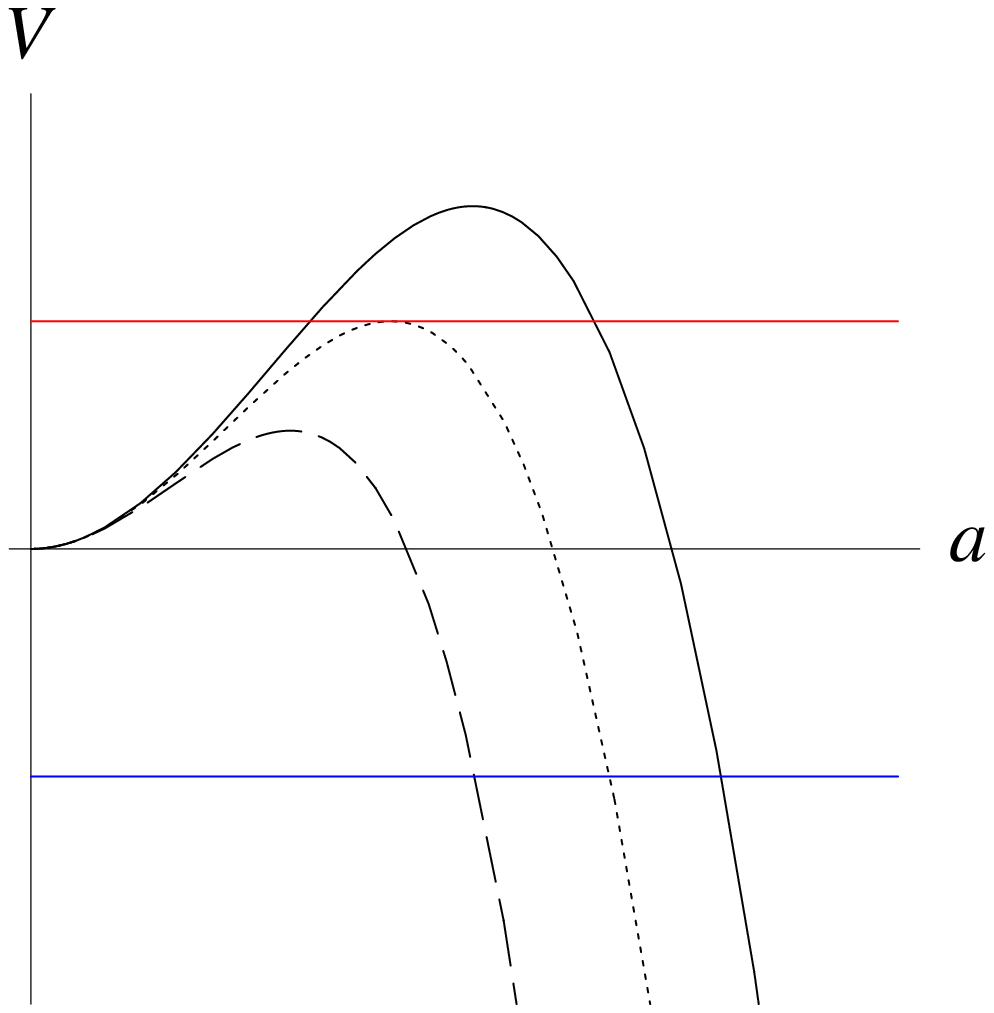}}  \vskip.5cm \scalebox{0.65}{\hskip1cm \includegraphics{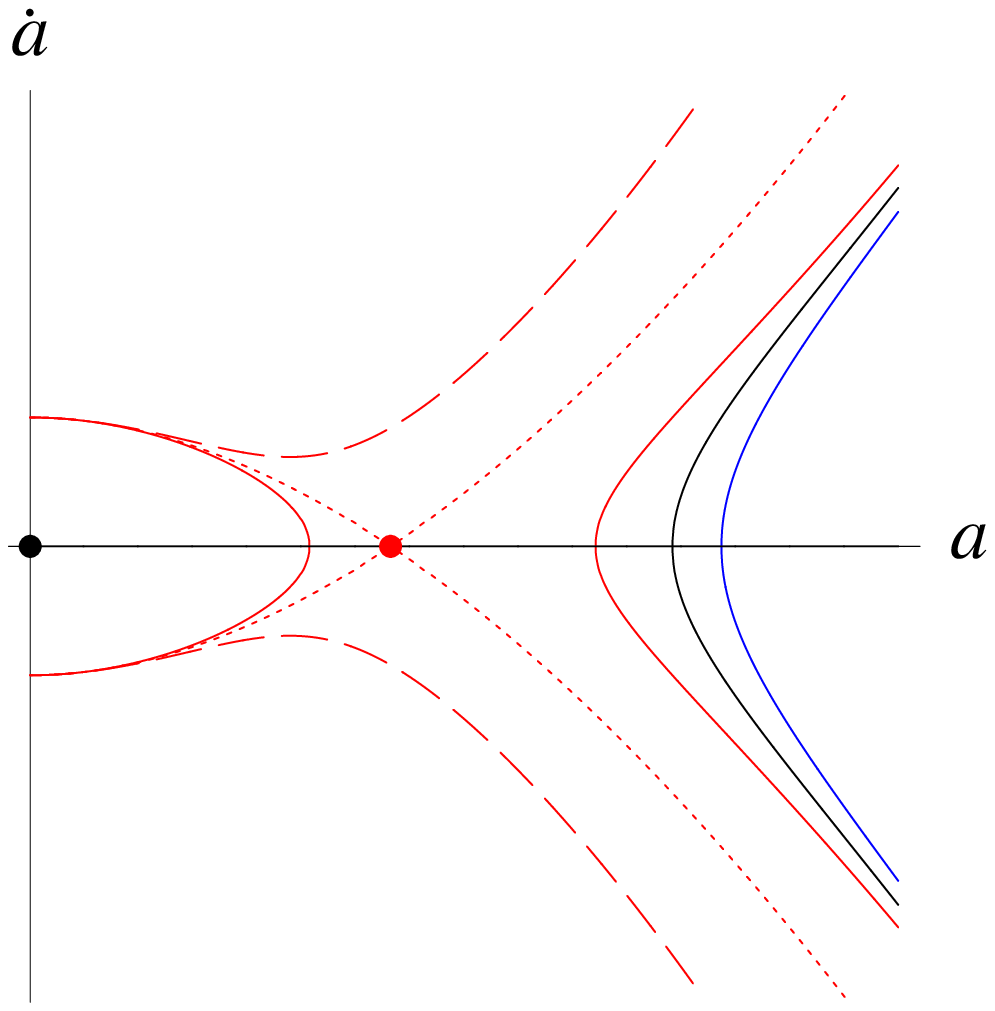}}
\hfil}
\vskip.25cm
\caption{\small Potential energy and phase portrait for $w<-1$, $\Lambda<0$, and three values of $A$.  The phase curves have been drawn for all the three values of $k$ only for the smaller value of $A$ (solid curve $V(a)$, solid phase curves).  For the other potential energies, only the phase curves corresponding to $k=-1$ are plotted.}
\label{F:lala}
\end{figure}

\begin{figure}[htbp]
\vbox{ \hfil
\scalebox{0.8}{\includegraphics{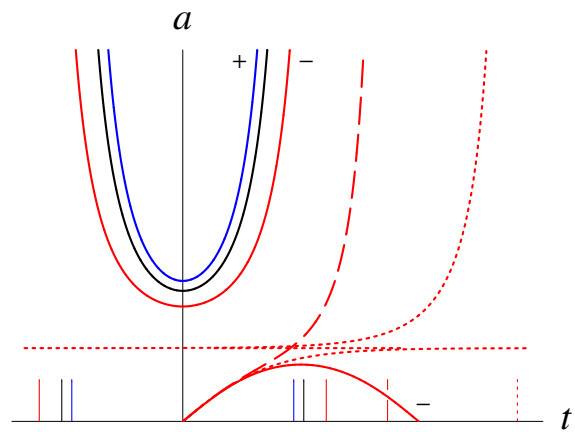}}
\hfil }
\vskip.25cm
\caption{\small Time dependence of the scale factor corresponding to the phase curves of Fig.~\ref{F:lala}.  Note the presence of vertical asymptotes where $a\to +\infty$ at a finite time.}
\label{F:param-bigripcasino}
\end{figure}

The potential energy and the phase portrait are shown in Fig.~\ref{F:lala}.  The time evolution of the scale factor is plotted in Fig.~\ref{F:param-bigripcasino}.  For $k=1$ and $k=0$, the universe starts with an ``anti-big rip'', contracts to a minimum size, and then re-expands symmetrically until a big rip occurs.  (Again, for $k=0$ there is also a static singular solution $a(t)\equiv 0$, corresponding to the dot at the origin of the phase plane.)  For the value~\eqref{bara} of the scale factor, the potential energy has a maximum, given by Eq.~\eqref{Vbar}.~\cite{newfootnote}  Hence, there are three possibilities for $k=-1$, which lead to qualitatively different behaviours.  

If $V(\bar{a})>c^2/2$, there are two disconnected ranges for the possible values of $a$, corresponding either to a big bang followed by a big crunch, or to a model in which $a$ contracts from infinity to a minimum size in a finite time and then re-expands symmetrically, as in the cases with $k=1$ and $k=0$.

If $V(\bar{a})=c^2/2$ we have a separatrix (dotted curves in Figs.~\ref{F:lala} and~\ref{F:param-bigripcasino}).  There is an unstable static solution $a(t)\equiv\bar{a}$, analogous to Einstein's static model of Sec.~\ref{S:Einstein}, represented by the dot with $a\neq 0$ in the phase portrait.  There are also a solution that starts with a big bang and approaches the static model asymptotically for $t\to +\infty$, and one that coasts the static model for an infinite time and then ends with a big rip (as well as their time-reversed versions).

Finally, if $V(\bar{a})<c^2/2$ (dashed curves), the universe starts with a big bang and ends with a big rip (or the time-reversed process).  Interestingly, during this evolution $\ddot{a}$ changes sign from negative to positive.  If $V(\bar{a})$ is very close to the critical value $c^2/2$, there is a phase in which $a(t)$ is approximately equal to the value $\bar{a}$ and increases very slowly, as in the Lema\^{\i}tre coasting models of Fig.~\ref{F:Lemaitre}.


\section{Final remarks}
\label{sec5}
\setcounter{equation}{0}

Using standard techniques of qualitative study, familiar from courses in classical mechanics, we have investigated the time evolution of Friedmann-Lema\^{\i}tre-Robertson-Walker models with a perfect fluid and a cosmological constant.  The analysis is elementary and does not require one to solve differential equations (except in some very simple cases, if one also wants to know the detailed time behaviour in some asymptotic regimes).  

There are several ways in which our discussion could be expanded.  Although we have considered only cases in which the parameter $w$ in the equation of state~\eqref{w} is fixed for all times, one can easily construct ``patchwork'' models where the cosmic evolution is dominated by different kinds of matter at different eras.  Also, one can combine the behaviour found for $a$ with Eq.~\eqref{rho-a} in order to find the time evolution of the energy density $\rho$.  Furthermore, the qualitative analysis can be straightforwardly extended to a generic barotropic equation of state~\eqref{state}.  Indeed,  integrating Eq.~\eqref{energy} gives
\begin{equation}
\int_{\rho_0}^\rho\frac{\mathrm{d}\rho}{\rho+p(\rho)}=\ln\left(\frac{a_0}{a}\right)^3,
\end{equation}
from which one finds $\rho$ as a function of $a$.  This can then be replaced into the Friedmann equation~\eqref{Friedmann} to obtain the potential energy for the equivalent mechanical system.  For example, a generalised Chaplygin gas with equation of state 
\begin{equation}
p=-C\,\rho^{-\alpha},
\end{equation}
with $C>0$ and $0<\alpha\leqslant 1$, has sometimes been considered as an alternative to phantom energy, in order to explain the cosmic acceleration.~\cite{moschella, bertolami}  This form of $p(\rho)$ leads to %
\begin{equation}
\rho=\left(C+Da^{-3(\alpha+1)}\right)^{\frac{1}{\alpha+1}},
\end{equation}
with $D=\left(\rho_0^{\alpha+1}-C\right)a_0^{3(\alpha+1)}$, and thus to a contribution
\begin{equation}
-\frac{4\pi G}{3 c^2}\left(C+\frac{D}{a^{3(\alpha+1)}}\right)^{\frac{1}{\alpha+1}}a^2
\end{equation}
to the potential energy $V(a)$ in Eq.~\eqref{totalenergy}.  It is again a simple exercise to find the asymptotic behaviours and to perform the qualitative analysis.

As an alternative to qualitative analysis, one could integrate numerically the Friedmann equation~\eqref{Friedmann'} to produce the plot for $a(t)$.  Although one might well prefer such a strategy in professional research work, we believe our approach to be pedagogically more valuable.  With the help of a phase portrait, a student knows at a glance the behaviour of a cosmological model under all possible initial conditions, recovering also cases that are likely to escape when one performs only a few numerical experiments.  (This happens, e.g., close to a separatrix; for example, it would be hard to hit numerically on the Lema\^{\i}tre models of Fig.~\ref{F:Lemaitre}, unless one knows already what to look for.)  In order to produce good-quality figures, all the plots in this article have indeed been obtained numerically.  However, it is important to realise that they could be reproduced with all their essential features, using only pencil and paper and a few techniques known from elementary calculus, so the technique presented here allows students to quickly and easily grasp the main features of a cosmological model.  

Of course, not all the combinations of $w$ and $\Lambda$ are equally important for an understanding of the standard model of cosmology.  Only the content of Sec.~\ref{Ss:13} ($w>-1/3$) is mandatory for an introductory course.  However, as we have seen, the generality of the method allows an instructor to easily present some conclusions pertaining to other, more exotic, values of $w$ whenever that might be useful.

The material presented in this paper can be used not only in a course on cosmology, but also as an unusual illustration of the qualitative techniques during lectures on classical dynamics, thus allowing an interesting interplay between concepts from different areas of physics.  Although the analysis follows the traditional scheme, the cosmological interpretation of the variable $a$, and the corresponding description in terms of the behaviour of models of the universe, should provide additional motivations for the students, who often find these subjects more fascinating and attractive than the point particles of classical mechanics.


\section*{Acknowledgements}

It is a pleasure to thank Valerio Faraoni for helpful comments.



\begin{thebibliography}{99}
\bibitem{fn1}
The treatment in Ref.~\onlinecite{uzan} does not allow one to obtain directly the behaviour of the scale factor, which is probably what most students are interested in.
\bibitem{fn5}
Note that now both $1+3w$ and $\Lambda$ are negative, and $1+w$ is positive, so $a_0$ given by Eq.~\eqref{Einsteinstatic} is a real number.
\bibitem{newfn}
Indices $a$ and $b$ denotes tensor components in four-dimensional spacetime, $\mathrm{\sl g}_{ab}$ is the metric, and $u_a=\mathrm{\sl g}_{ab} u^b$, where $u^a$ is the fluid four-velocity.
\bibitem{conservation}
Of course, the property $\rho=\mathrm{const}$ also follows, when $w=-1$, from Eq.~\eqref{energy}, to which  the condition $\nabla^b T_{ab}=0$ reduces for the FLRW metric~\eqref{metric}.
\bibitem{newfootnote}
Note that now both $1+3w$, $1+w$, and $\Lambda$ are negative, so $\bar{a}$ given by Eq.~\eqref{bara} is a real number, and $V(\bar{a})$ is positive.
\end{thebibliography}
\end{document}